\documentclass[preprint,prb,aps,showpacs]{revtex4}
\usepackage{graphicx} 
\usepackage{dcolumn} 
\usepackage{bm} 
\usepackage{amsmath,amssymb}

\begin{document}

\title{On the Kondo effect in carbon nanotubes at half halfing}

\author{B.\ Babi{\'c}}
\author{T.\ Kontos}
\author{C. Sch{\"o}nenberger}
\email{Christian.Schoenenberger@unibas.ch}
\homepage{www.unibas.ch/phys-meso}
\affiliation{Institut f\"ur Physik, Universit\"at Basel, Klingelbergstr.~82,
    CH-4056 Basel, Switzerland }
\date{\today}

\begin{abstract}

In a single state of a quantum dot the Kondo effect arises
due to the spin-degeneracy, which is present if the dot is occupied
with one electron ($N=1$). The eigenstates of a carbon nanotube quantum dot
possess an additional orbital degeneracy leading to a four-fold
shell pattern. This additional degeneracy increases the
possibility for the Kondo effect to appear.
We revisit the Kondo problem in metallic carbon nanotubes by
linear and non-linear transport measurement in this regime, in
which the four-fold pattern is present. We have analyzed the
ground state of CNTs, which were grown by chemical vapor deposition,
at fillings $N=1$, $N=2$, and $N=3$.
Of particular interest is the half-filled shell, i.e. $N=2$.
In this case, the ground state is either a paired electron state
or a state for which the singlet and triplet states are effectively degenerate,
allowing in the latter case for the appearance of the Kondo effect.
We deduce numbers for the effective missmatch $\delta$ of the levels from
perfect degeneracy and the exchange energy $J$. While $\delta \sim 0.1-0.2$
(in units of level spacing) is in agreement with previous work, the
exchange term is found to be surprisingly small: $J\alt 0.02$.
In addition we report on the observation of gaps, which in one
case is seen at $N=3$ and in another is present over an extended
sequence of levels.
\end{abstract}

\maketitle


\section{Introduction}

In the past decade, transport measurements have emerged as a
primary tool for exploring the electrical properties of
structures on the nanometer scale.
Due to their unique electronic bandstructure, much attention has been focused on carbon
nanotubes (CNTs).\cite{Dresselhaus1996} For metallic single wall carbon nanotubes
(SWNTs) just two spin degenerate one-dimensional ($1$d) modes should
govern their transport properties at low energies, which makes
them interesting model systems to explore the physics in reduced
dimensions.\cite{Dekker-PhysToday-1999}

Due to the finite length, given by the lithographically fabricated
contacts on opposite sides of the CNT (two-terminal device with source and drain contacts),
the one-dimensional CNT is turned into a quantum dot~\cite{Tans-1998} at
low temperatures (typically at \mbox{$\alt 10$\,K}), i.e. into a zero-dimensional
object with a discrete level spectrum. The confinement is formed by the finite back-reflection
at the edges of the contacts. The level spacing \mbox{$\delta E$}
is determined by the contact separation \mbox{$L$} and is inversely proportional to it.
This particle in the box-model holds provided the level-broadening $\Gamma$ and
the temperature are both smaller than $\delta E$. $\Gamma$ describes
the life-time broadening proportional to the coupling strength to the leads.

Until now, three transport regimes have been identified in SWNTs:
A) single-electron tunneling,\cite{CB-Reviews} which is dominated
by the on-site Coulomb repulsion expressed by the energy term
\mbox{$U=e^2/C_{\Sigma}$}, where $C_{\Sigma}$ is the total
capacitance; B) the regime of correlated transport, in which
higher-order tunneling processes, are appreciable, leading to the emergence
of the Kondo
effect;\cite{Goldhaber-Gordon-1998,Kouwenhoven-Glazman-2001} and
C) the open SWNT for which Coulomb interaction may be neglected
and the residual gate-dependence of $G$ can be described as in a
tunable Fabry-Perot resonator.\cite{Fabry-Perot-2001} A) holds
for low, B) for intermediate and C) for high transparent contacts.
The Kondo-effect, which occurs at intermediate contact
transparency, can be seen as the Holly Grail of many electron
physics. It has first been observed in quantum dots by
Goldhaber-Gordon {\it et al.}~\cite{Goldhaber-Gordon-1998} and in
CNTs by Nyg{\aa}rd {\it et al.}\cite{Nygard-2000} In contrast to
the Coulomb Blockade (CB) regime, which only probes the electrons
confined on a QD, the Kondo effect incorporates delocalized
electrons in the leads coherently. The presence of a degenerate
ground state in the quantum dot (for example, a singly occupied
level with two-fold degeneracy due to the spin degree of freedom)
forms the basis for the Kondo effect. A multitude of coherent
second- and higher-order elastic tunneling processes between the
Fermi seas of the leads and the quantum-dot state are enabled,
leading to the appearance of a narrow peak in the density of
states (DOS) right at the Fermi level (the Kondo resonance) at
sufficiently low temperatures. Its width is given by the Kondo
temperature $T_K$ which measures the binding energy of this
many-electron state. The peak in the DOS enhances the probability
for electrons to tunnel from source to drain. As a consequence the
zero-temperature linear-conductance saturates at the quantized
conductance \mbox{$G_0=2e^2/h$} (unitary limit), provided the
device is coupled symmetrically to source and drain.

As in atoms, eigenstates in quantum dots (QDs) may be degenerate due to symmetries
and together with the spin degeneracy and Pauli principle lead to the formation
of electronic shells. Indeed, striking shell patterns have been observed in
QDs.\cite{Q-Dots-1984,Tarucha-1996} The eigenstates (Bloch-states) at the Fermi energy
of graphene (two-dimensional sheet of graphite) is two fold degenerate. The two
wave functions correspond to the two carbon sublattices (the unit-cell is composed of two C-atoms).
This degeneracy is preserved in CNTs and should therefore lead to two degenerate orbitals
in a finite-length nanotube in $0$d. Together with the spin degeneracy, the shells are
expected to be four-fold degenerate. This shell pattern has
recently been observed by Buitelaar and coworkers \cite{Buitelaar-2002} in multi-wall carbon nanotubes (MWNTs)
and by Liang {\it et al.} in SWNTs.\cite{Liang-2002}
Within one shell the ground-state spin was shown to follow the sequence
\mbox{$S=0\rightarrow 1/2\rightarrow 0 \rightarrow1/2$} in the former work,
whereas a possible triplet ground state for two added electrons was suggested by the latter
authors, i.e. the sequence \mbox{$S=0 \rightarrow 1/2 \rightarrow 1 \rightarrow 1/2$}.

Here we focus on CVD-grown metallic carbon nanotubes
(CNTs).\cite{Chemical-Vapor-Deposition,BabicKirchberg2004} We
will first examine the four-fold shell pattern in great detail and
demonstrate that the half-filled ground state (i.e. $2$ electrons
added to an empty shell) is either a paired electron with $S=0$ or
the six possible two-electron states are effectively degenerate due to a level
broadening exceeding the orbital mismatch and exchange energy. We
furthermore have discovered striking gaps in several samples. This
anomaly is at present not understood.

\section{Experimental}

Single wall carbon nanotubes (SWNTs) have been grown from patterned
catalyst islands by the chemical vapor deposition method on
\mbox{Si/SiO$_2$} substrates.\cite{Babic2003} The degenerately doped silicon,
terminated by a \mbox{$400$\,nm} thick \mbox{SiO$_2$} layer, is used
as a back-gate to modulate the electrochemical potential of the
SWNT electrically contacted with a source and drain terminal.
The contacts are patterned by electron-beam
lithography (EBL) using polymethylmethacrylate (PMMA) as resist,
followed by metallization with palladium and lift-off.\cite{Dai-Pd-2003,BabicKirchberg2004}
Once the samples are made, semiconducting and metallic
SWNTs are distinguished by the dependence of their electrical
conductance $G$ on the gate voltage $V_g$ measured at room
temperature (\mbox{$T \approx 300$\,K}).\cite{BabicKirchberg2004}
In the rest of the paper we report on measurements performed on metallic SWNTs with
relatively low-ohmic contacts, such that co-tunneling and Kondo
physics is observable.

The electrical characterization of the devices has been
performed at low temperature in a \mbox{$^3$He} system at \mbox{$300$\,mK}.
We measure the electrical current $I$ with a low noise current
to voltage amplifier as a function of source-drain \mbox{($V_{sd}$)} and gate \mbox{($V_g$)}
voltage and determine the differential conductance  \mbox{$G_d:=\partial I/\partial V_{sd}$}
numerically. Finally, the collected data $G_d(V_{sd},V_g)$
are represented in a two-dimensional grey-scale representation in which the grey-scale
corresponds to the magnitude of $G_d$. The linear-response conductance
$G:=I/V_{sd}$ with $V_{sd} \rightarrow 0$ is measured at a small but finite
source-drain voltage of \mbox{$40$\,$\mu$V}.

\section{Results and Discussion}

In this section, we will focus first on one set of measurements
which we will analyze in great detail. This set of data is shown
in Fig.~1. Fig.~1a shows the linear-response conductance $G$ as a
function of gate voltage $V_g$. Fig.~1b and 1c display the
corresponding grey-scale plots of the differential conductance
$G_d$ in zero magnetic field and \mbox{$B=5$\,T}, respectively.
White corresponds to low and black to high conductance.

The observed patterns correspond to a quantum dot with a
relatively strong coupling to the contacts. Signatures of the
latter are high conductance `ridges', observed at zero bias
\mbox{($V_{sd}\approx 0$)} and $B=0$, caused by the the Kondo
effect. This effect is a many-electron effect and requires a
relatively high tunneling coupling to the leads in order to be
appreciable at temperatures where the measurements take place. As
required,\cite{Goldhaber-Gordon-1998,Kouwenhoven-Glazman-2001} we
find that $G$ increases if the temperature is lowered below
\mbox{$\approx 4$\,K} to saturate at the lowest temperature close
to the unitary limit of $G=2e^2/h$. The characteristic energy
scale, i.e. the Kondo temperature $T_K$, has been deduced from the
temperature dependence of $G$ in ridge $3$ (not shown) and found
to be \mbox{$T_K\approx 2$\,K}. The conductance enhancement due to
the Kondo effect is observed at zero source-drain voltage if
\mbox{$B=0$}. In a magnetic field, however, the conductance
enhancement is reduced and a splitting of the peak conductance to
finite source-drain voltages is
expected.\cite{Goldhaber-Gordon-1998,Kondo-in-field} This
splitting is visible in Fig.~1c which was measured in a perpendicular magnetic field of
\mbox{$5$\,T}. That the linear-response conductance $G$ is
suppressed in a magnetic field is clearly seen in Fig.~1a in which
the solid (dashed) curve correspond to $B=0$ \mbox{($B=5$\,T)}.

Because the many-electron effects (Kondo effect) are suppressed in
magnetic field, we can use the linear-response conductance
measurement in magnetic field (dashed curve in Fig.~1a) to assign
the charge states of the quantum dot with reference to the
single-electron tunneling picture. A transition from a ground
state with $N$ electrons in the dot to one with $N+1$ gives rise
to a peak in the conductance, whereas $G$ is suppressed in
between. This pattern is nicely seen in the dashed curve of
Fig.~1a, in which transitions have been labelled. Evidently, these
conductance peaks form a repeating pattern in clusters of four.
This pattern is the generic shell pattern of an ideal CNT quantum
dot.\cite{Buitelaar-2002,Liang-2002} It is caused by the
four-fold degeneracy of $0$d-eigenstates, two of which stem from
spin and the other two from the so-called \mbox{$K-K^{\prime}$}
orbital-degeneracy of graphene.\cite{Dresselhaus1996} The
four-fold pattern can be regarded as a measure of the quality of
the SWNTs. It is not observed in all SWNTs and even if observed it
is not usually present over the whole gate voltage range. But it
can repeat over several shells, not just two as shown in Fig.~1.
The degeneracy may be lifted by disorder and by the contacts which
may couple differently to the two orbital states. As has been
pointed out be Oreg~{\it et al.}, the four fold pattern may be
absent even in a `perfect' SWNT because the two orbital states can
respond differently to the electrostatic gate-field if
inhomogeneous.\cite{Oreg-2000}

Let us continue to analyze
our data in terms of the constant-interaction model.\cite{constant-interaction-model}
In order to assign the states only two parameters are needed: the single-electron charging
energy $U:=e^2/C_{\Sigma}$ (which can be expressed by the total
capacitance $C_{\Sigma}$ and is assumed to be a constant in this model)
and the single-electron level spacing \mbox{$\delta E$}. Note, that $\delta E$ measures the
energy difference between a filled shell to a state with one additional
electron. This is sketched in the inset of Fig.~1a. To add an electron
one has to provide an `addition energy' composed of charging energy $U$ plus
level-spacing $\delta E$, the latter only if the electron must be added into
a new shell. Therefore, the addition energy $\Delta E$ equals $\delta E + U$ for the
first electron in a shell, whereas it amounts to `only' $U$ for the following
three added electrons belonging to the same shell.
Since $\Delta E$ is proportional to the
gate-voltage difference between adjacent conductance peaks
(the conversion factor equals $eC_g/C_{\Sigma}$),
the labelling of states in terms of charge $N$ in Fig.~1a should be understandable.
$N=0~mod~4$ corresponds to a ground state with a filled shell.
Due to the large addition energy, the conductance is strongly suppressed
for a filled shell, giving rise to the diamond-like white areas (denoted by A, B and C) in
the grey-scale plots. Adding electrons to the filled shell one by one (peaks
in $G$, dashed curve of Fig.~1a), we reach
the state $N=4$ which corresponds again to a filled shell.

In the following, the
ground states will be labelled by $N=0...3$ for the first quartet and
$N=0^{\prime}...3^{\prime}$ for the second, where $N=4=0^{\prime}$.
Relying on the constant interaction model,
the ratio between the average level spacing and charging energy
amounts to $\delta E/U \approx 1$ in our data. It is seen, however, that $U$ is constant
to a good approximation,
but that $\delta E$ varies. For the respective diamonds A, B and C, the
level spacing $\delta E$
amounts to $\approx 7$, $5$, and \mbox{$3$\,meV}, respectively. Theoretically, the
level-spacing of an ideal SWNT is given by \mbox{$\delta E=hv_F/2L$},
where \mbox{$v_F = 8\times 10^5$\,m/s} is the Fermi velocity and $L$ the length of
the tube that determines the $1d$ cavity.\cite{Dresselhaus1996} Taking the
nanotube length $L$ measured from the edges of the contacts, which for this
sample amounts to \mbox{$L\sim 300$\,nm}, the equation predicts a level-spacing
of \mbox{$\delta E \sim 5.5$\,meV} in good agreement with the experimental values
of \mbox{$3-7$\,meV}. The data in Fig.~1 yields for the charging energy
\mbox{$U=5.3\pm 0.5$\,meV} and a gate-conversion factor of \mbox{$\alpha:=C_g/C_{\Sigma}$}
of $0.08$.

Focusing on the high-conductance Kondo ridges at zero bias voltage, we see in
Fig.~1b a ridge at charge states $N=1$ and $N=3$, whereas $G$ is suppressed at
half-filling, i.e. at $N=2$. The situation is different for the second quartet, where
Kondo ridges are observed for all three states $N=1$, $2$ and $3$. This phenomenon
was reported before by Liang {\it et al.}\cite{Liang-2002} Whereas a spin-$1/2$ Kondo
effect is expected for $N=1$ (one electron) and $N=3$ (one hole),
the situation at half-filling, i.e. at $N=2$ is less obvious. The observed Kondo-effect
was assigned to a spin-$1$ triplet state in Ref.~\cite{Liang-2002}. In the following we
re-examine this assignment. To do so, we have to go beyond the `free' electron model
and consider among other things the exchange interaction. There are three additional
parameters: First, it has been pointed out that the orbital degeneracy need
not to be exact.\cite{Buitelaar-2002,Liang-2002} The orbital mismatch is denoted by $\delta$.
With regard to on-site charging energy the doubly occupancy of one orbital
is a bit higher in energy as compared to placing each of the two electrons
in a separate orbital. This parameters has been introduced by Oreg {\it et al.}~\cite{Oreg-2000}
and is denoted by $\delta U$. Finally, placing the two electrons in different orbitals
gives rise to a spin-dependent exchange energy term, which according to Hund's rule
favors the triplet state, i.e. the state with spin $S=1$. This parameter is denoted by
$J$. These parameters have been extracted, both for MWNTs~\cite{Buitelaar-2002} and
SWNTs~\cite{Liang-2002} and the analysis of our data confirm the previously obtained values.
The importance of the
parameters in descending order is $\delta$, $J$ and $\delta U$ as the least important one.
The former work by Buitelaar {\it et al.} reports $\delta \approx 0.2$ and $J < 0.09$ and the
latter work by Liang {\it et al.} reports $\delta \approx 0.3$, $J \approx 0.1$ and
$\delta U < 0.1$ (all numbers are measured in units of level spacing $\delta E$).
We neglect $\delta U$ because it is small and typically much smaller than the bare level broadening
$\Gamma$, which - as we will emphasize - matters as well.
Since the Kondo effect is the dynamic screening of the local spin by exchange with a sea of
electrons, it is tempting to assign the Kondo ridge for $N=2$ to a spin-$1$ (triplet) ground
state. However, with regard to the just mentioned parameters, this appears to be unlikely, because
$J<\delta$, favoring a spin-$0$ ground state.\cite{Oreg-2000}

Let us first contrast the possible states at $N=1$ and $N=2$ for
the case of degenerate orbitals  \mbox{($\delta=0$)} and without
exchange \mbox{($J=0$)}, shown in Fig.~2a, with the case of a
finite level mismatch and a finite exchange energy, shown in
Fig.~2b. In the first (maybe too naive) model of Fig.~2a, the
degeneracy equals $4$ at $N=1$ and clearly Kondo physics can
emerge. At $N=2$ the degeneracy is even larger, amounting to $6$
and second-order elastic spin-flip processes are energetically
allowed so that Kondo physics can emerge as well. Here, two states
are paired-electron states and the other four may be labelled as
one singlet state with spin $S=0$ and three triplet states with
$S=1$, denoted as S and T states. The Kondo effect may be expected
to be even enhanced in this case due to the larger number of
states.\cite{Kondo-Kouwenhoven-2000} This scenario corresponds to
the Kondo effect for which the singlet and triplet states are
degenerate. This has been realized experimentally in
semiconducting quantum dots by tuning the states with either
a magnetic or an electric field.\cite{Kondo-Kouwenhoven-2000,Ensslin-PRL-2003, Nazarov1}
Once we go over to the more realistic model shown in Fig.~2b, assuming
that exchange and level mismatch are non-zero and of comparable
magnitude, the $N=1$ states remains `normal' in the sense that
only the lowest lying orbital need to be considered. At
half-filling, i.e. at $N=2$, there are however two
possibilities:\cite{Oreg-2000} if exchange dominates
\mbox{($J>\delta$)}, the ground state is the spin triplet (T)
state, whereas if the opposite holds, the ground state is a paired
electron (PE) residing on the lowest orbital state. The energy
difference between the T and PE states is given by
\mbox{$\Delta=\delta-J$}.\cite{Pustilnik-2000} Although the Kondo
effect is (in principle) possible for the triplet, there is no
Kondo effect possible for paired electrons. Note, that unlike
previous discussions, there are {\em three} cases to consider at
half-filling. Two may give rise to Kondo and one does not. To have
an abbreviation at hand we denote the three $N=2$ states with ST
(degenerate ground state), PE (paired electron ground state) and T
(triplet ground state). As mentioned above, the Kondo ridge at
$N=2$ has been assigned to the triplet state.\cite{Liang-2002}
Though this is tempting at first sight, there is no necessity. In
fact, this assignment is unlikely, first because $J$ is measured
to be small (usually smaller than $\delta$) and secondly, the
Kondo temperature $T_K$ for triplet Kondo is predicted to be much
smaller than $T_K$ for $S=1/2$-Kondo~\cite{Pustilnik-2001} (an
estimate will follow below).

We now look at the excitation spectra for $N=1,3$ and $N=2$ in zero field. This is shown in
Fig.~3 (a-c), where (a) and (b) correspond to odd filling ($N=1,3$)
and (b) to half-filling ($N=2$). The energy $\Delta$ of the first
excited state at fixed $N$ relative to the ground state is given by the level mismatch
$\delta$, both for $N=1$ and $N=3$, i.e. $\Delta_{1,3}=\delta$. This is illustrated in the respective
insets on the right. The first excited states show up as a conductance peak at finite $V_{sd}$,
corresponding to the excitation energy. This is a so-called inelastic co-tunneling process.
We obtain from the measurement \mbox{$\Delta_1=0.92$\,meV} and \mbox{$\Delta_3=0.85$\,meV}.
Hence, the level mismatch is given by \mbox{$0.89\pm 0.4$\,meV}.
For the $N=2$ case we have to distinguish two possibilities: if $J > \delta$, the ground state is the triplet (T)
and the excited state the paired electron (PE) state, yielding \mbox{$\Delta_2=J-\delta$}. If on the other
hand $J < \delta$, the states are reversed, yielding \mbox{$\Delta_2=\delta - J$}. In general,
\mbox{$\Delta_2=|\delta-J|$}. From the experiment (Fig.~1c) we deduce \mbox{$\Delta_2=0.88$\,meV}.
We stress that we measure on one and the same shell so that we can use the parameter $\delta$, measured
for the $N=1,3$ case, also for the $N=2$ case. Comparing the numbers, leaves open two possibilities:
either the exchange parameter is quite small, i.e. $J\sim 0$
(taking the possible errors into account \mbox{$J \alt 0.1$\,meV}), or it is quite large
\mbox{$J \agt 1.65$\,meV}. If the latter would be true, it would be a remarkable coincidence
that we find $|\delta-J|\sim \delta$ with $J\approx 2\delta$. Moreover, the ratio
$J/\delta E > 0.3$ would be quite remarkable with regard to previous measurements.
On the other hand, comparable values for $J$ have theoretically been predicted, however only for small diameter tubes.
For example, $J/\delta E$ was estimated to be $\geq 0.22$ and
$\geq 0.44$ for a $(10,10)$ and $(5,5)$ tube, respectively.\cite{Oreg-2000}
However, the diameter $d$ of CVD-grown NTs is known to vary substantial and in particular
we find that \mbox{$d\agt 2$\,nm},\cite{BabicKirchberg2004} from which one would
theoretically predict an exchange parameter of order $J/\delta E \sim 0.1$, which disagrees
with the finding above.
If $J$ were indeed as large as \mbox{$2\delta$}, and therefore $J>\delta$,
the triplet state would be the ground state at half-filling, i.e. at
$N=2$ and $N=2^{\prime}$. The Kondo effect at half-filling must then be assigned to
the $S=1$ Kondo effect. To explain the absence of the Kondo effect for $N=2$ and its
presence for $N=2^{\prime}$, one would have to argue that the Kondo temperature is
smaller than \mbox{$300$\,mK} at $N=2$, whereas it is larger at $N=2^{\prime}$.
Pustilnik \textit{et al.} \cite{Pustilnik-2003}
showed that the Kondo temperature \mbox{$T_K^{S=1}$}
for the triplet state is smaller than \mbox{$T_{K,1/2}$} for the spin-$1/2$ case.
More precisely, $T_{K,1}$ can be estimated according to
\mbox{$k_B T_{K,1}=\left(k_B T_{K,1/2}\right)^2/\delta E$}.
The average width of the zero-bias resonances at $N=1,3$ in the left quartet is measured to be
\mbox{$0.35$\,meV}, yielding as a prediction \mbox{$T_{K,1}\approx 0.25$\,K}.
In the right quartet the same procedure yields for $N=1^{\prime},3^{\prime}$
a mean width of \mbox{$0.8$\,meV}, from which one predicts \mbox{$T_{K,1}\approx 1.5$\,K}.
Hence, the comparison with the measuring temperature does not exclude
$S=1$ Kondo, as \mbox{$T_{K,1}\alt 0.3$\,K} in the left quartet
and \mbox{$T_{K,1} > 0.3$\,K} in the right one. However, the ratio $T_{K,1}/T_{K,1/2}$
measured in the right quartet is inconsistent with theory which predicts
$T_{K,1/2}/\delta E$. The former is evaluated to $0.8\dots 1.6$, whereas the latter is
at most $0.3$. In simple terms, the appearance of the two resonances
at $N=1^{\prime}$ and $N=2^{\prime}$ with essentially one and the same width
($1.1$ and $0.9$\,meV, respectively), makes triplet-Kondo quite unlikely.

In magnetic field the states further split due to the Zeeman
energy given by $\pm g\mu_B B/2$, where $\mu_B$ is the Bohr
magneton and $g$ the so-called $g$-factor. $g$ has been measured
in related electrical measurements on carbon nanotube quantum dots
and found to agree with the free electron value of $g=2$.\cite{Buitelaar-2002,Tans-1998,McEuen-1998}
Due to the Zeeman-splitting the excitation spectrum changes. At $N=1$ and for
small magnetic fields (with regard to the level mismatch), the
spin-$1/2$ Kondo resonance is expected to
split,\cite{Buitelaar-2002} evolving into inelastic co-tunneling
with an excitation energy given by \mbox{$\Delta_Z=g\mu_B B$}.
Because of the relatively large width of the zero-bias resonances,
this shift is hardly visible for small magnetic fields in the
experiment. That is why we have chosen a relatively large field of
\mbox{$5$\,T}. This field yields for the Zemann excitation energy
\mbox{$0.58$\,meV}, taking $g=2$. Note, however, that there is a
second excited state given by the level mismatch \mbox{$\delta
\approx 0.9$\,meV}. If we analyze the non-linear differential
conductance as a function of $V_{sd}$, we see two excitation lines
(one at positive and one at negative bias), which are markedly
broadened, suggesting an overlap of two excitation features, see
Fig.~3a (grey overlaid graph). The onset of the excitation peaks
agrees with the Zeeman energy (arrows from below). This analysis
is of particular importance in the $N=2$ case, because it allows
us to distinguish the PE from the T ground state unambiguously,
see Fig.~4. If the ground state is the paired-electron state, the
first excitation occurs at energy $\Delta_2=\delta-\Delta_Z -J$
$\approx \delta-\Delta_Z$ (because $J\approx 0$) and the second
lies at $\delta$. In contrast, if the ground state is the spin-$1$
triplet state the first two excited states have energy $\Delta=J+
\Delta_Z-\delta$ $\approx \delta + \Delta_Z$ and $J+\Delta_Z$
$\approx 2\delta + \Delta_Z$ (because $J\approx 2\delta$ in this
case). This is shown (approximately to scale) in the illustrations
of Fig.~4b and c, respectively. Based on the field-dependence of
the excitation spectrum we can predict the position of the
excitation peaks for the two cases. In the measurement, shown in
Fig.~4a, the upper black arrows point to expected excitations if
the ground state is the paired-electron (PE) state, whereas the
upwards pointing open arrows correspond to the expected
excitations for the triplet (T) ground state. It is obvious that
the agreement with the PE state is much better. The excitation
peaks at zero field do not move out to larger energies expected
for the T ground state, but rather shrink. In particular there are
clear low-energy shoulders visible which agree quite reasonably
with the expected lowest energy excitation energy for the PE
ground state.

Taking all arguments together, this makes a convincing case for
the ground state at half-filling, i.e. for $N=2$, which is the
paired-electron state. Moreover, the exchange energy must be very
small. How do we then have to explain the pronounced Kondo ridge
at $N=2^{\prime}$, visible in Fig.~1b ? As we have pointed out
when discussing Fig.~2, there are two cases at half-filling that
allow for Kondo: spin-$1$ Kondo in case of the triplet state or
the degenerate ground state, i.e. the ST-state. Based on our
previous discussion the former can be excluded, so that the only
remaining possibility requires degenerate orbitals. We know that
the orbitals are not exactly degenerate. The level mismatch, as
deduced from the $N=1-3$ states, amounts to \mbox{$\delta\approx
0.9$\,meV}, which is quite appreciable. Due to the relative large
width of the zero-bias conductance peaks at $N=1^{\prime}\dots
3^{\prime}$ we are not able to deduce the level mismatch on the
second shell along the same lines as before for the first shell.
Though it is possible that
$\delta$ is smaller in the second shell, it is unlikely zero.
We emphasize, however, that it is crucial to compare the level mismatch with the level width,
due to the tunneling couplings $\Gamma_s$ and $\Gamma_d$ to the respective contacts.
If $\delta < \Gamma$, where $\Gamma:=\Gamma_s + \Gamma_d$, the two orbital states cannot be distinguished and
are in effect degenerate. We know from other measurements on carbon nanotubes
that $\Gamma$ may vary a lot with gate voltage. Our picture of the
half-filled state is correct, if we can show that
$\Gamma$ is smaller than $\delta$ within the first shell, but larger within the second.
There are several ways to deduce $\Gamma$. One possibility is to deduce it from the width of
the excitation features at fixed $N$, another one is to analyze the transitions at finite bias at the
edge of the Coulomb-blockade (CB) diamonds. For the left shell the excitation spectra for states
$N=2$ and $N=3$ (see Fig.~3) yield \mbox{$\Gamma\approx 0.7$\,meV}, whereas a cut at \mbox{$V_g=3.1$\,V},
corresponding to the transition $0\leftrightarrow 1$, yields \mbox{$\Gamma\approx 0.9$\,meV}.
Because we cannot resolve excitation features in the right shell, we have to rely on
transitions at the edge of CB-diamonds. We deduce \mbox{$\Gamma\approx 3$\,meV} at
$0^{\prime}\leftrightarrow 1^{\prime}$ and \mbox{$\Gamma\approx 1.9$\,meV} at
$3^{\prime}\leftrightarrow 4^{\prime}$. Clearly $\Gamma \alt \delta$ for the left shell and
$\Gamma > \delta$ for the right one in support of our statement. To conclude this part, we can say that
the Kondo effect at $N=2^{\prime}$ is not a triplet Kondo,
but arises because $\Gamma$ is larger than the level mismatch,
resulting in a ground state in which the paired-electron, the singlet and
triplet states are effectively degenerate. Our data is only consistent with a very small
exchange exchange term of \mbox{$J/\delta E \alt 0.02$}. Such a small value can only be reconciled
with theory~\cite{Oreg-2000} if either the tube has a large diameter
of order \mbox{$\sim 10$\,nm} or the interaction is locally screened, possibly by the presence of
other nanotubes forming a bundle.

Examination of the measured data shows that for the Kondo
resonances labelled with $1'$ and $3'$ in Fig.~1b, the positions
of the maximum conductance are situated at non-zero bias. This is
shown in Fig.~5. This phenomenon has been observed in
semiconducting quantum dots and was termed the anomalous Kondo
effect by Simmel {\it et al.}\cite{Simmel-1999} It was suggested
by these authors that the effect is due to asymmetric and
energy-dependent coupling strengths $\Gamma_s$ and $\Gamma_d$ to
the two reservoirs. The effect has thereafter been confirmed
theoretically in a single-impurity Anderson
model.\cite{Krawiec2002} The authors show that the peak
conductance is shifted provided that $\Gamma_s \not = \Gamma_d$,
but an energy dependence of $\Gamma_{s,d}$ is not required. We
stress here, however, that the Anderson model introduces an
additional model-dependent asymmetry in that
$U\rightarrow \infty$, which is not realized in a real quantum
dot. At half-filling, there is particle-hole symmetry where
electrons (holes) can be exchanged via both the bare state at
energy $\epsilon_0$ and the one at $\epsilon_0+U$. In this case,
no shift of the Kondo peak is expected even if $\Gamma_s \not =
\Gamma_d$. Extrapolating $G(T)$ to the unitary limit $G(0)$ at
zero temperature (not shown) using the standard expression to fit
the Kondo effect, i.e.
$G(T)=G(0)/\left[1+(2^{1/s}-1)(T/T_K)^2\right]^2$,\cite{Kondo-fit}
we obtain for the ridge at charge state $N=3$ a zero temperature
conductance of \mbox{$G(0)=1.68$\,$e^2/h$}, out of which the
$\Gamma$ ratio is estimated to be $\approx 2$. Hence, there is an
asymmetry of magnitude comparable to Ref.~\cite{Simmel-1999}. Our
statement, that the shift of the Kondo peak to finite bias is
absent at half-filling is beautifully reflected in the data of
Fig.~5. Due to the four-fold symmetry, half-filling corresponds to
charge state $N=2^{\prime}$ and indeed, this peak has its maximum
at \mbox{$V_{sd}=0$}. The other two resonances are shifted
oppositely, one to $V_{sd}>0$ \mbox{($N=1^{\prime}$)} and the
other to $V_{sd}< 0$ \mbox{($N=3^{\prime}$)}. The shift amounts to
\mbox{$0.22$\,meV}. These shifts are comparable in magnitude to
the ones seen by Schimmel~{\it et al.}, although they have
observed only unipolar shifts. Finally we remark that the
transitions to the Coulomb-blockade diamonds, i.e.
$2^{\prime}\leftrightarrow 1^{\prime}$ and
$3^{\prime}\leftrightarrow 4^{\prime}$ are asymmetric with respect
to the $V_{sd}$, see arrows. Cross-sections at constant
gate-voltage through these transitions allow to deduce the
respective $\Gamma$'s and their ratio:
$\gamma:=\Gamma_s/\Gamma_d$. We point out, that this asymmetry is
a consequence of the level degeneracy. Consider tunneling at
finite bias into the $N=1^{\prime}$ state. Because there is a
four-fold degeneracy the effective in-tunneling rate is enhanced
by a factor of $4$. In contrast, this phase-space argument does
not hold for the out-tunneling rate. The respective current steps
are then given by $(4e/h)\Gamma_s \Gamma_d/(\Gamma_s + 4\Gamma_d)$
for one bias polarity (e.g. \mbox{$V_{sd}>0$}) and $(4e/h)\Gamma_s
\Gamma_d/(4\Gamma_s + \Gamma_d)$ for the other polarity, where the
factor $4$ counts the degeneracy. It is clear from these two
relations that the current steps are only different for the two
polarities if $\gamma\not = 1$. The two current steps, measured
for the transition \mbox{$3^{\prime}\leftrightarrow 4^{\prime}$},
amount to \mbox{$\approx 20$} and \mbox{$\approx 30$\,nA},
yielding for the $\Gamma$-ratio \mbox{$\gamma\sim 2$} (in
agreement to what we have deduced before in a different way) and
\mbox{$\Gamma_s\approx 1.4$\,meV} and \mbox{$\Gamma_d\approx
0.7$\,meV}, so that the total level broadening is approximately
\mbox{$\Gamma\approx 2$\,meV}. Also the latter value is in
agreement with the previously mentioned width of the transition,
which we measured to be \mbox{$\Gamma\approx 1.9$\,meV}.

After this extensive analysis we use the last part of this section to
point to observed deviations. Fig.~6 displays the dependance of the
linear-response (a) and differential conductance (b) of another sample
also contacted with Pd. The contact separation is longer and
amounts to \mbox{$L\sim 0.8$\,$\mu$m}.
The linear-response conductance is bound
by \mbox{$2e^2/h$} suggesting that we measure through
one individual SWNT. Four-fold clustering in the electron
addition spectrum is observed for more then five shells ($A-E$),
corresponding to $20$ electrons. Due to the three
times larger length of this device as compared to the one in Fig.~1
the energy scale is reduced by approximately a factor of three.
The level-spacing amounts to \mbox{$\delta E\approx 1-1.5$\,meV} and
the charging energy to \mbox{$U\approx 1$\,meV}. The ratio
$\delta E/U \approx 1$ as before.
As with the data of Fig.~1 the
Kondo effect may appear at half-filling \mbox{($\alpha$)} or
may be absent \mbox{($\beta$)}, which according to the discussion
above would correspond to the PE and S ground state, respectively.
There are differences, however. The most dramatic one occurs in shell $D$
for the three electron state (filling $3/4$), marked with $O$.
Instead of the expected spin-$1/2$ Kondo, the conductance is
actually suppressed. This is seen as a pronounced white bubble.
Because the Kondo effect is present for the one electron state
(filling $1/4$), this implies breaking of particle-hole symmetry.
This effect is quite surprising and has not been reported before.
We do not have a convincing explanation but mention one possibility.
The three electrons at $N=3$ may like to form
a high-spin state with total spin \mbox{$S=3/2$}. However, this requires
three different orbitals, but there are only two in an ideal tube.
It may be that the nanotube is not perfect, rather a bundle or a multi-shell
tube, which may provide additional orbitals. We think that this
scenario is unlikely, because we just have shown that the exchange
is small, and it is particularly small
if the interaction is screened by other tubes.
The gap may however be induced by a magnetic defect caused by
residual catalyst particles, which may enhance the exchange energy.
Due to the small size of catalyst particles such a defect interacts
only locally. If efficient, one would expect a strong effect on the addition
energy due to the large energy scale of the defect. In our opinion,
the observed regularity of the addition pattern rules out disorder.

Similar gap-features are sometimes seen over the entire grey-scale
plot. We show in Fig.~7 a short section taken out of an extensive
differential conductance grey-scale plot of another sample. The contacting material
is Au in this case and the contact separation amounts to \mbox{$L\sim 1$\,$\mu$m}. The contact
transparencies are lower here and typical two terminal conductances
are of order \mbox{$0.1$\,$e^2/h$}. Consequently, the main
features in the differential-conductance are Coulomb blockade (CB) diamonds.
The generic four-fold shell structure is not apparent. It is masked
by the charging energy which dominates here. The observed addition energy
amounts to \mbox{$\Delta E \approx 5$\,meV}.
We stress that the $dI/dV_{sd}$ measurements of Fig.~7a extend
over more than $17$ electrons without any noticeably change. The
linear-response conductance (Fig.~7c) shows a very regular set of
high conductance peaks at the transition between neighboring
charge states with peak values approaching \mbox{$0.8$\,$e^2/h$}.
The spacing between these CB-oscillation peaks is surprisingly
constant, amounting to \mbox{$\Delta V_g = 73 \pm 5$\,mV}. This
yields a gate-coupling constant of \mbox{$\alpha = 0.068$}, which
is comparable to the one deduced for the sample of Fig.~1
\mbox{($\alpha = 0.08$)}.

We present this measurement here, because of the presence
of a striking gap-structure, which is seen inside of {\em all} CB diamonds
and which might be related to the gap which we have mentioned
before, i.e. the feature labelled $O$ in Fig.~6b.
Two $dI/dV_{sd}$ cross-sections at constant $V_g$ are presented in Fig.~7b.
We find that the size of the gap \mbox{$\Delta_{g}$} varies a bit
in different charge state and is estimated
to be \mbox{$\Delta_g\approx 0.7$\,meV} \mbox{($0.3\dots 0.9$\,meV)}.
Additional suppression may be caused, if the nanotube is split by a
strong scattering center into two segments in series. In this case, however,
a regular periodic CB-oscillation pattern is not expected, because
single-electron transport requires that two charge states are degenerate
in both segments simultaneously. While this maybe possible occasionally,
it would be surprising if the levels would move in both segments with gate voltage
exactly equally. We therefore are convinced, that this scenario is wrong.
Also a possible parallel conductance through two (or more) different tubes can be
excluded, because this should appear in the grey-scale-plot as a bare superposition of two
CB-patterns. Moreover, the observed grey-scale plot cannot be modelled as a regular
CB-pattern multiplied by a gap-feature in the vicinity of \mbox{$V_{sd}\approx 0$}.
This is evident from the Fig.~7c which shows $G(V_g)$ at \mbox{$V_{sd}=0$} (full curve) and
at \mbox{$V_{sd}=1.5$\,mV} (dashed curve). In the shaded region, corresponding to Fig.~7a,
the suppression is only active in between the CB-oscillation peaks, whereas the peaks
themselves are not suppressed, suggesting that the $0$d orbitals extend from source
to drain. The low-conductance `bubbles' are therefore confined
to the CB-region of the nanotube and this new effect is
observable in transport through a single carbon nanotube. This does not
mean that there is only one single-walled carbon nanotube present. The device may still
consist of a small bundle or a multishell tube, of which only one tube is
effectively coupled to the reservoirs. In addition, the presence of magnetic
impurities in the form of catalyst particles cannot be excluded, so that the observed gaps
may originate from magnetic interactions with these particles. The Kondo effect which
results in a high conductance resonance can be described as an anti-ferromagnetic
exchange between the leads and the quantum dot. It is tempting to suggest that the
opposite scenario, namely ferromagnetic exchange with, for example, a magnetic particle,
may suppress the conductance.\cite{Fiete2002}

\section{Conclusion}
In conclusion, we have analyzed the ground-state of carbon nanotubes which are
relatively strongly coupled to the attached leads. Spin $1/2$ Kondo is present
for charge states $N=1$ and $N=3$. At half-filling, i.e. for two electrons on the
dot, the ground-state is either a non-degenerate paired electron or a highly degenerate
two-electron state. Whereas the Kondo effect is prohibited in the first case, it is
allowed (and enhanced) in the second. The appearance of the Kondo effect at $N=2$
is largely determined by the magnitude of the level broadening $\Gamma$, caused by
the coupling to the leads. We have also observed striking gaps whose origin need to
be unravelled in the future.

\begin{acknowledgments}
We acknowledge contributions and discussions to this work by W.
Belzig, V. N. Golovach, D. Loss, L. Forr\'o (EPFL).
Support by the Swiss National Science Foundation, the NCCR on
Nanoscience and the BBW (Cost and RTN DIENOW) is gratefully acknowledged.
\end{acknowledgments}


\newpage

\begin{figure}[htb!]
\begin{center}
DUE TO SIZE LIMITATIONS OF THE COND-MAT SERVER, THE PAPER INCLUDING FIIGURES CAN BE
DOWNLOADED FROM: www.unibas.ch/phys-meso/Research/Papers/2004/Kondo-4shell-SWNT.pdf
\end{center}
\caption{\label{Fig1} (a) Linear response conductance $G$ as a
function of back-gate voltage $V_g$ of a SWNT device with contact
separation \mbox{$L \sim 300$\,nm} (edge-to-edge of reservoirs),
measured at \mbox{$T$=300\,mK} and in a magnetic field of
\mbox{$B=0$} (solid curve) and \mbox{$B=5$\,T} (dashed curve). A
clear clustering in four peaks is observed (pronounced in magnetic
field), which suggests a single-electron shell pattern with
four-fold degeneracy. Charge states corresponding to a filled
shell (inset) are labelled as $0$ or $4$. (b,c) Corresponding
grey-scale plots of the differential conductance
\mbox{$dI/dV_{sd}$} (darker more conductive) at \mbox{$B=0$} (b)
and \mbox{$B=5$\,T} (c) as a function of gate \mbox{$V_g$} and
source-drain voltage \mbox{$V_{sd}$}. In the first shell, high
conductance Kondo ridges (visible at \mbox{$V_{sd}\sim 0$}) are
observed for charge states $1$ and $3$, whereas they appear for
states $1^{\prime}$, $2^{\prime}$, and $3^{\prime}$ in the second
shell. The Kondo ridges clearly split in the applied magnetic
field.}
\end{figure}

\begin{figure}[htb!]
\begin{center}
DUE TO SIZE LIMITATIONS OF THE COND-MAT SERVER, THE PAPER INCLUDING FIIGURES CAN BE
DOWNLOADED FROM: www.unibas.ch/phys-meso/Research/Papers/2004/Kondo-4shell-SWNT.pdf
\end{center}
\caption{\label{Fig2}
Illustration of the state-filling scheme for
one \mbox{($N=1$)} and two \mbox{($N=2$)} excess electrons.
In (a) the level-mismatch $\delta$ and the exchange energy $J$
are zero, whereas these parameters are non-zero in (b). PE denotes
a paired-electron state, S (T) the singlet (triplet) two electron state.
The Kondo-effect may arise in three cases: obviously for the spin-$1/2$
with one excess electron ($N=1$) and if $N=2$ for the spin-$1$ triplet state, but
also for the case for which $\delta=J=0$, i.e. when the singlet and triplet states
are degenerate (ST state).}
\end{figure}

\begin{figure}[htb!]
\begin{center}
DUE TO SIZE LIMITATIONS OF THE COND-MAT SERVER, THE PAPER INCLUDING FIIGURES CAN BE
DOWNLOADED FROM: www.unibas.ch/phys-meso/Research/Papers/2004/Kondo-4shell-SWNT.pdf
\end{center}
\caption{\label{Fig3}
Non-linear differential conductance $dI/dV_{sd}$ as a function of $V_{sd}$ at
\mbox{$V_g=const$} deduced from the data shown in Fig.~1b at zero magnetic field.
(a) and (b) correspond to the states $N=1$ and $N=3$, whereas (c) correspond to
the half-filled shell $N=2$. All three $dI/dV_{sd}$ cuts have been placed in the
middle of the charge state. The visible excitation peaks occur at energy
$\Delta$ and are due to inelastic co-tunneling through the excited state. The
relevant states are illustrated in the respective insets on the right.
The grey curve in (a) has been measured in a magnetic field of \mbox{$5$\,T}. Arrows
point to \mbox{$\Delta_Z=g\mu_B B=0.58$\,meV} using $g=2$.
}
\end{figure}

\begin{figure}[htb!]
\begin{center}
DUE TO SIZE LIMITATIONS OF THE COND-MAT SERVER, THE PAPER INCLUDING FIIGURES CAN BE
DOWNLOADED FROM: www.unibas.ch/phys-meso/Research/Papers/2004/Kondo-4shell-SWNT.pdf
\end{center}
\caption{\label{Fig4}
(a) Non-linear differential conductance $dI/dV_{sd}$ as a function of $V_{sd}$ taken
from the data shown in Fig.~1b at a fixed gate-voltage corresponding to $N=2$. The
thick (thin) curve was measured in a magnetic field of \mbox{$B=5$\,T} \mbox{($B=0$\,T)}.
(b) and (c) illustrate the magnetic-field dependence of the first two excited states
at $N=2$. The two cases are drawn approximately to scale using the fact that
$J$ is either $\approx 0$ (PE ground state) or $\approx 2\delta$ (T ground state)
deduced from the data of Fig.~1 and Fig.~3.}
\end{figure}

\begin{figure}[htb!]
\begin{center}
DUE TO SIZE LIMITATIONS OF THE COND-MAT SERVER, THE PAPER INCLUDING FIIGURES CAN BE
DOWNLOADED FROM: www.unibas.ch/phys-meso/Research/Papers/2004/Kondo-4shell-SWNT.pdf
\end{center}
\caption{\label{Fig5}
The Kondo resonance is observed to be offset with respect to the bias voltage
\mbox{$V_{sd}$} for the mixed-valence state with filling $1/4$ \mbox{($N=1^{\prime}$)}
and $3/4$ \mbox{($N=3^{\prime}$)}, whereas it is centered at \mbox{$V_{sd}=0$} at
half-filling. (a) shows the respective differential conductance at constant gate voltage
corresponding to part (b), which reproduces the second shell of Fig.~1b. Arrows emphasize
an additional asymmetry, discussed in the text.}
\end{figure}

\begin{figure}[htb!]
\begin{center}
DUE TO SIZE LIMITATIONS OF THE COND-MAT SERVER, THE PAPER INCLUDING FIIGURES CAN BE
DOWNLOADED FROM: www.unibas.ch/phys-meso/Research/Papers/2004/Kondo-4shell-SWNT.pdf
\end{center}
\caption{\label{Fig6} (a) Linear response conductance plotted as a
function of the gate voltage $V_g$ and (b) differential
conductance \mbox{$dI/dV_{sd}$} (darker more conductive) plotted
as a function of \mbox{$V_g$} and \mbox{$V_{sd}$} for an another
SWNT device with length \mbox{$L\sim 800$\,nm} contacted by
palladium. The shell pattern of four electrons each extends over
$5$ shells ($A-E$). The Kondo effect occurring at half-filling is
marked with $\alpha$, while $\beta$ corresponds to the singlet
ground state. $O$ points to an anomaly, a strong gap-feature
arising for a three electron state. The non-linear $dI/dV_{sd}$
through the middle of this state is shown in (c).}
\end{figure}

\begin{figure}[htb!]
\begin{center}
DUE TO SIZE LIMITATIONS OF THE COND-MAT SERVER, THE PAPER INCLUDING FIIGURES CAN BE
DOWNLOADED FROM: www.unibas.ch/phys-meso/Research/Papers/2004/Kondo-4shell-SWNT.pdf
\end{center}
\caption{\label{Fig7}  (a) Differential-conductance plot of a SWNT
device with contact separation \mbox{$L \sim 1$\,$\mu$m} at
\mbox{$T$=300\,mK} (maximum conductance \mbox{$=e^2/h$}, black).
Coulomb blockade diamonds are clearly seen. In addition gaps
appear near zero bias. (b) \mbox{$dI/dV_{sd}$} as a function of
$V_{sd}$ and at constant $V_g$ along the corresponding lines,
shown in (a). Conductance as a function of gate voltage taken at
\mbox{$V_{sd}=0$} (full) and \mbox{$V_{sd}=1.5$\,mV} (dashed). The
shaded region correspond to (a).}
\end{figure}

\end{document}